\documentclass[12pt]{iopart}
\usepackage{iopams}

\usepackage[]{graphicx}
\usepackage[T1]{fontenc}
\usepackage{dcolumn}% Align table columns on decimal point
\usepackage{sistyle}
\usepackage{color}

\begin{document}

\title{An absolute Johnson noise thermometer}   % This is the title of the article
\author{Luca Callegaro\textsuperscript{1},
		Vincenzo D'Elia\textsuperscript{1},
        Marco Pisani\textsuperscript{1}, and
        Alessio Pollarolo\textsuperscript{1} \\[\medskipamount]  % Authors lines
        {\small \textsuperscript{1}INRIM - Istituto Nazionale di Ricerca Metrologica} \\
        {\small Strada delle Cacce, 91 - 10135 Torino, Italy}}    % Authors addresses

\begin{abstract}
We developed an absolute Johnson noise thermometer (JNT), an instrument to measure the thermodynamic temperature of a sensing resistor, with traceability to voltage, resistance and frequency quantities. The temperature is measured in energy units, and can be converted to SI units (kelvin) with the accepted value of the Boltzmann constant $k_\mathrm{B}$; or, conversely, it can be employed to perform measurements at the triple point of water, and obtain a determination of $k_\mathrm{B}$. The thermometer is composed of a correlation spectrum analyzer an a calibrated noise source, both constructed around commercial mixed-signal boards. The calibrator generates a pseudorandom noise, by digital synthesis and amplitude scaling with inductive voltage dividers; the signal spectrum is a frequency comb covering the measurement bandwidth. JNT measurements at room temperature are compatible with those of a standard platinum resistance thermometer within the combined uncertainty of \SI{60}{\micro K  \, K^{-1}}. A path towards future improvements of JNT accuracy is also sketched.

\end{abstract}

\maketitle
\section{Introduction}

The accurate measurement of Johnson noise has been considered a method for determining the thermodynamic temperature, and the Boltzmann constant $k_\mathrm{B}$, since its very first observation \cite{Johnson28}. A resistor $R$, in thermodynamic equilibrium, generates a noise voltage $v(t)$ with the spectral power density\footnote{The expression is accurate to one part in $10^7$ at room temperature and frequency below \SI{1}{MHz}.} $S^2_v =4 R \mathcal{T}$, where $\mathcal{T}$ is its thermodynamic temperature measured in energy units.

Within the International System of units, to the quantity temperature $T$ a base unit is associated, the kelvin (\SI{}{K}), defined by assigning the temperature of the triple point of water (TPW), $T_\mathrm{TPW} = $ \SI{273.16}{K}; the relation $\mathcal{T_\mathrm{TPW}} = k_\mathrm{B} T_\mathrm{TPW}$ defines $k_\mathrm{B}$. In an effort towards a possible redefinition of the kelvin, in 2005 the Consultative Committee for Thermometry (CCT) of the International Committee for Weights and Measures (CIPM) \cite{CCT2005} recommended to "initiate and continue experiments to determine values of thermodynamic temperature and the Boltzmann constant". 

Johnson noise thermometry experiments have the potential for such new determinations. A detailed analysis suggests \cite{Storm86} the possibility of achieving a $k_\mathrm{B}$ relative uncertainty of a few parts in $10^6$, comparable to estimated or forecasted uncertainties of other existing or proposed experiments \cite{Fellmuth2006}. 

In the following, we present a Johnson noise thermometer (JNT) which measures $\mathcal{T}$ with traceability to national standards of ac voltage, resistance and frequency. If $k_\mathrm{B}$ is taken as given (in the following the CODATA 2006 adjustment \cite{CODATA2006_RevModPhys, CODATA2006_JPhysChem} will be employed) the JNT measurement outcome can be compared with an ITS-90 temperature $T_{90}$ measurement taken as reference.

Presently, the JNT has been tested by performing measurements near room temperature. A measurement run at room temperature gives $T$ in in agreement with $T_{90}$ within the combined relative measurement uncertainty around \SI{60}{\micro K  \, K^{-1}}. Several uncertainty contributions are related to the use of commercial instrumentation. In the future, purposely-built instruments under development will permit a significant accuracy improvement, and will open the possibility of employing the JNT for a new determination of $k_\mathrm{B}$.

\section{Absolute and relative measurements}

The main difficulty in the development of an accurate JNT is the faintness of the Johnson noise, which must be amplified by a large factor ($10^4 - 10^6$). Noise added by front-end amplifiers has an amplitude comparable with that of the Johnson noise itself, but can be rejected with the correlation technique (see Ref. \cite[par. 6.4]{White96} for a review, and \cite{White2008b} for an extended mathematical treatment of digital correlation). An adequate rejection requires a careful design of the correlator, and in particular of its front-end amplifiers \cite{White84, Storm89, White2000}. The drawback of most effective amplifier designs criteria is a poor stability of gain and frequency response. Hence, an automated in-line gain calibration subsystem has to be incorporated in the JNT.

We may call \emph{relative} JNTs \cite{White96} those where the calibration signal is also given by Johnson noise of a resistor (the same, or a different one), placed at a known reference temperature, which is typically $T_\mathrm{TPW}$. A relative JNT measures the ratio $\mathcal{T}/\mathcal{T}_\mathrm{TPW} = T/T_\mathrm{TPW}$. 

An \emph{absolute} JNT measures $\mathcal{T}$ directly; therefore, the calibration signal has to be traceable to electromechanical SI units. Although in the past \cite{Storm86} a detailed proposal of an absolute JNT has been published, the only absolute JNT now in operation is at the National Institute of Standards and Technology (NIST) \cite{Nam2003,White2008,Benz2009}. In NIST's JNT the calibration signal is generated by a synthesized pseudorandom voltage noise source based on pulse-driven Josephson junction arrays. The calibration signal amplitude is thus directly linked to the Josephson fundamental constant $K_\mathrm{J}$, and the driving frequency of the Josephson array.

In the setup here presented, the calibration signal is generated with a commercial digital-to-analog (DAC) board, measured with an ac rms thermal voltmeter, and scaled in amplitude with a chain of electromagnetic voltage dividers (see Sec. \ref{sec:calsig}).

\section{Overview of the implementation}

\begin{figure}[ht]
	\centering
	\includegraphics[width=4in]{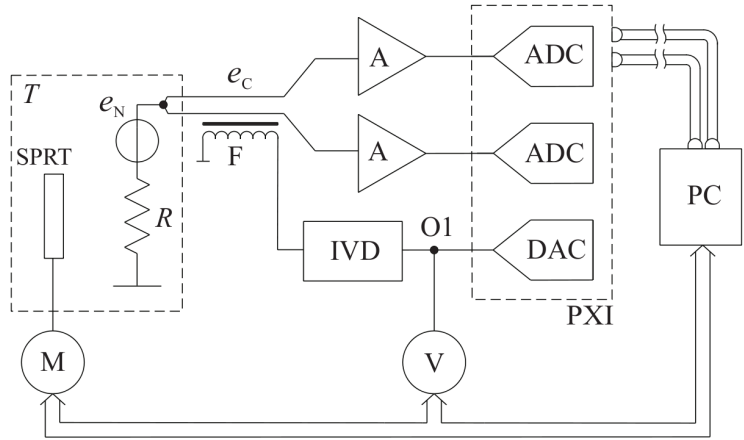}	
	\caption{Block schematics of the JNT, see text for details.}
	\label{fig:schemablocchi}
\end{figure}

The block schematics of the JNT is shown in Fig. \ref{fig:schemablocchi}. 

The probe resistor $R$, generating the Johnson noise $e_\mathrm{N}$, is at temperature $T$, measured with a standard platinum resistance thermometer SPRT and a resistance meter M. The signal $e_\mathrm{N}$ is acquired by two identical acquisition channels in parallel, each one composed of an amplifier A and an analog-to-digital converter ADC. Resulting digital codes are transmitted by an optical fibre interface to the processing computer PC, which implements a digital correlation spectrum analyzer algorithm.  

The signal $e_\mathrm{C}$ is employed to periodically calibrate the analyzer gain, and injected in series with $e_\mathrm{N}$. $e_\mathrm{C}$ is a pseudorandom noise, with the same bandwidth $B$ of the measurement. It is generated by PC and a digital-to-analog converter DAC (also connected with the optical link). The waveform is measured by a voltmeter V, and reduced in amplitude by inductive voltage dividers IVD and an injection feedthrough transformer F. The measurements of V and M are acquired by PC through an IEEE-488 interface bus. 

\section{Details of the implementation}

\subsection{Probe resistor}
The probe is a single resistor $R$, enclosed in a cylindrical screen and connected to the amplifiers A by a shielded four-wire cable. Presently a Vishay mod. VSR thick film resistor, having a nominal value of \SI{1}{k\ohm}, is employed. $R$ is calibrated in dc regime, but a relative frequency deviation lower than \SI{5E-6}{} up to \SI{10}{kHz} is expected \cite{Bohacek2006}.

\subsection{Amplifiers}
Amplifiers A are identical; each one is composed by two stages, giving an overall gain of $\approx$\SI{31000}{}. The first stage (see Ref. \cite{Callegaro08112866} for details) is a pseudo-differential, cascode FET amplifier in an open-loop configuration \cite{White2000}. Its equivalent input voltage noise is $\approx$\SI{0.8}{nV/\sqrt{Hz}}, and the bias current is $\approx$\SI{2-3}{pA}; the gain flatness is better than \SI{1}{dB} over \SI{1}{MHz} bandwidth. It is battery-powered (by separate battery packs for each channel to avoid residual correlations due to limited power-supply rejection ratio), and electrostatically shielded. Both first stages are placed in in a $\mu$metal box which acts as a magnetic shield. A second conventional op amp stage, working also as a rough bandpass filter and having with a separate power supply, follows. 

\subsection{Spectrum analyzer}
The ADCs are those of a commercial board (National Instruments mod. 4462: \SI{24}{bit} resolution, \SI{204.8}{kHz} maximum sampling frequency, synchronous sampling of the channels), embedded in a PXI rack. Codes are acquired continuosly and transmitted by an optical fibre link interface (National Instruments mod. 8336) to PC, where a digital cross-correlation algorithm based on fast-Fourier transform computes and averages spectra $C(f_k)$ having up to $2^{17}$ discrete frequency points $f_k$. Acquisition and computation are performed in parallel to optimize the measurement time.

\subsection{Calibration}
\label{sec:calsig}

\begin{figure}[ht]
	\centering
	\includegraphics[width=5.5in]{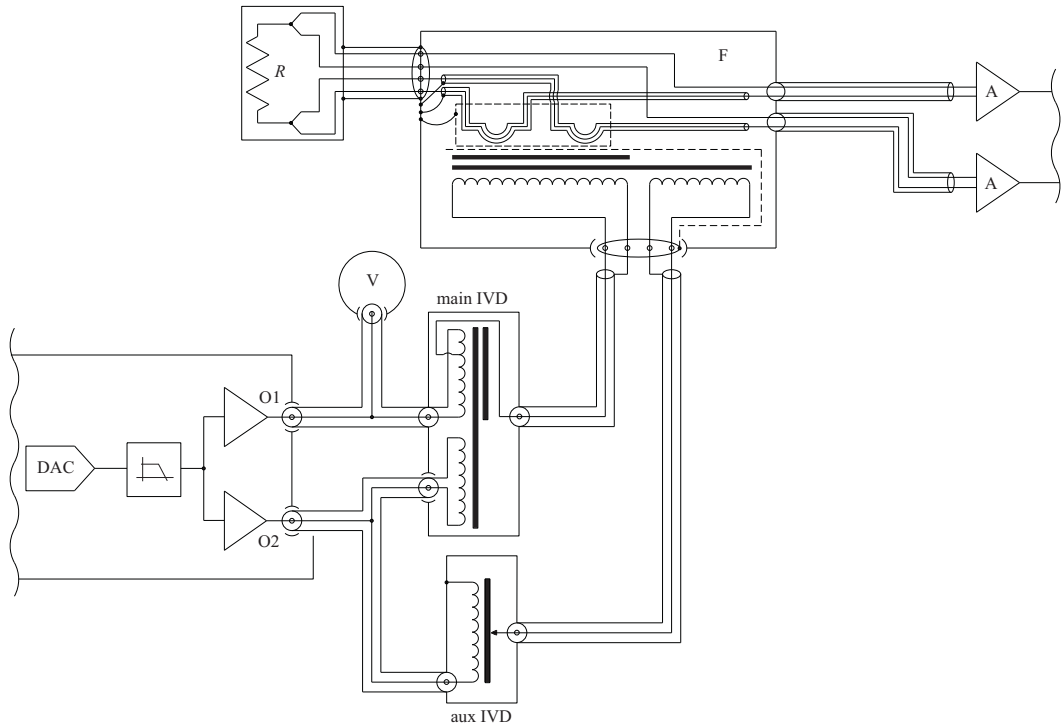}	
	\caption{Detailed schematics of the calibration subsystem, see text for details.}
	\label{fig:schemacal}
\end{figure}

The calibration subsystem is shown in Fig. \ref{fig:schemacal}. 

A commercial DAC board (National Instrument mod. 6733 board, \SI{16}{bit} resolution, \SI{1}{MHz} maximum sampling frequency, synchronized with the spectrum analyzer ADCs and embedded in the same PXI rack) generates the calibration waveform, which goes through an anti-alias filter (a 6-pole analog Butterworth lowpass filter with \SI{30}{kHz} bandwidth), and is buffered by two amplifiers with outputs O1 and O2 (each amplifier is provided with dc output restoration circuit \cite{AB008} to drive electromagnetic devices).

The signal is generated at an amplitude of $\approx$\SI{300}{mV}. One of the outputs (O1) is measured with a calibrated thermal voltmeter V: presently, a Fluke mod. 8506A voltmeter is employed. Because of its limited accuracy and stability, the 8506A reading is compared, immediately before and after each experiment, with the reading of a Fluke mod. 5790A ac measurement standard, traceable to the national standard of ac voltage\footnote{Unfortunately, the 5790A has a periodically fluctuating input impedance \cite{Simonson96} which causes glitches in the acquisition, and cannot be directly employed during the measurement.}.

After generation, the calibration waveform is scaled in amplitude (presently, by a factor of \SI{12000}{}) by two divider stages in cascade:
\begin{itemize}
\item a block of two inductive voltage dividers (IVD), both set at \SI{0.01}{} ratio. The main IVD is a fixed two-stage divider; its magnetizing winding is energized by O2, and the ratio winding by O1 (the same measured by V). The auxiliary IVD is a commercial 7-decade divider (Electro Scientific Instruments ESI mod. DT72) also energized by O2;
\item a two-stage feedthrough transformer F. Its magnetizing winding is energized by the auxiliary IVD output, while the primary winding by the main IVD output. Both windings have 120 turns. Two single-turn secondary windings (wound with opposite polarities) inject the calibration signal $e_\mathrm{C}$ in series with each connection of $R$ to the amplifiers. The two-stage technique not only improves the ratio accuracy but also reduces the burden of F on the main IVD. 
\end{itemize}
Both main IVD and F ratios are presently not calibrated, and the nominal turn ratio is employed in data processing.

\begin{figure}[ht]
	\centering
	\includegraphics[width=4in]{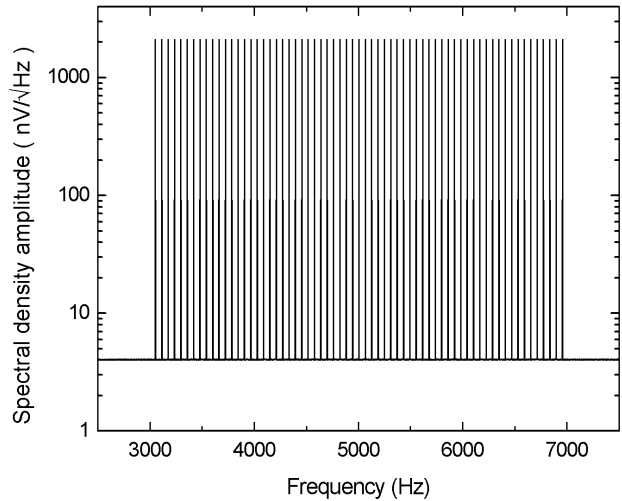}	
	\caption{The calibration signal $e_\mathrm{C}$ for a bandwidth $B=$\SI{3-7}{kHz}. The baseline is the Johnson noise $e_\mathrm{N}$. }
	\label{fig:comb}
\end{figure}

The calibration signal $e_\mathrm{C}$ is a pseudorandom noise, continuously recycled, having the power spectrum of an uniform frequency comb (see Fig. \ref{fig:comb}). The acquisition time for a single spectrum analyzer sweep is a multiple of the repetition period of the calibration signal, and since ADC and DAC sampling rates are commensurate and derived from the same clock, no spectral leakage occurs in the spectrum analyzer output.

At variance with NIST calibration waveform, whose comb covers the entire bandpass of the analyzer, $e_\mathrm{C}$ is band-limited: the comb frequencies cover uniformly the chosen measurement bandwidth $B$. The sinewave amplitudes are carefully adjusted in order to have a flat comb (relative deviations from the average amplitude better than \SI{1E-3}{}) at the output of the main IVD. Since the sinewaves have a relative phase chosen at random, $e_\mathrm{C}$  has an approximate Gaussian distribution of amplitudes.

\subsection{Thermometry}

Resistor $R$ is at the moment not thermostated, but simply kept in an isothermal equalization block at room temperature. Temperature $T_{90}$ is measured with a \SI{100}{\ohm} standard platinum resistance thermometer (SPRT), a Minco mod. S1060-2, calibrated  at the fixed points of the ITS-90, as maintained at INRIM \cite{Marcarino2003}. The meter M measuring the SPRT is presently a Agilent Tech. mod 3458A, option 002, whose calibration is traceable to Italian standard of dc resistance.

\section{Measurement procedure}

The measurement consists of $n$ repeated cycles, labeled $j=1...n$, each one composed of two phases: 
\begin{itemize}
\item measurement of Johnson noise spectrum $N^{j}(f_k)$ (the average of $m_\mathrm{N}$ sweeps). Temperature $T_{90}^j$ is also measured during this phase;
\item measurement of calibration spectrum $C^{j}(f_k)$ (the average of $m_\mathrm{C}$ sweeps). Calibration voltage $V^j$ is also measured during this phase.
\end{itemize}

In the present setup (and, correspondingly, Eq. \ref{eq:model}) we chose to inject $e_\mathrm{C}$ in series with $e_\mathrm{N}$ during the calibration phase, at variance with the standard method of alternately measuring $e_\mathrm{C}$ and $e_\mathrm{N}$. The advantage is a simplification in the layout, since no low-signal switching device is required, and electro-magnetic interferences (EMI) are easier to be kept under control. Moreover, impedance matching between the two phases is automatically achieved. On the other hand, the effect of possible spectrum analyzer nonlinearities has to be carefully considered \cite{White2008}.

The measurement model is:

\begin{eqnarray}
\label{eq:model}
	G^j = \frac{1}{V^j} \left\{ \Delta f \sum_{f_k \in B} \left[ C^j(f_k) - N^j(f_k) \right] \right\}^{1/2}, \nonumber \\
	\mathcal{T}^j =  \frac{1}{4 R \left( G^j \right)^2}  \sum_{f_k \in B} N^j(f_k), \nonumber \\
	\Delta T^j = k_\mathrm{B}^{-1} \mathcal{T}^j - T_{90}^j, \qquad \delta^j = \frac{\Delta T^j}{T_{90}^j}; 
\end{eqnarray}

where $G^j$ is the equivalent gain of the cross-correlator during cycle $j$ (average over the bandwidth $B$ of the calibration signal); and $\Delta f = f_{k+1} - f_{k}$ is the frequency bin corresponding to each $f_k$. $\Delta T^j$ is the deviation of JNT reading $\mathcal{T}^j$  with respect to the reference temperature $T_{90}^j$, and $\delta^j$ the same deviation expressed in relative terms.

\section{Results}

As an example of results, the following refers to a continuous acquisition run (the parameters of the acquisition are: $n=1200$ cycles, each of $2^{17}$ points at \SI{200}{kHz} sampling frequency, $m_\mathrm{N}=200$, $m_\mathrm{C}=50$, $B=$\SI{3-7}{kHz}; the calibration signal has a length of $2^{14}$ codes with \SI{500}{kHz} sampling rate). Total measurement time is \SI{2.3E5}{s} (about \SI{2.7}{d}).   % Fig. \ref{fig:errore_tempo} shows the relative deviation $\delta^j$ as a function of acquisition time $t$. 

\begin{figure}[ht]
	\centering
	\includegraphics[width=4in]{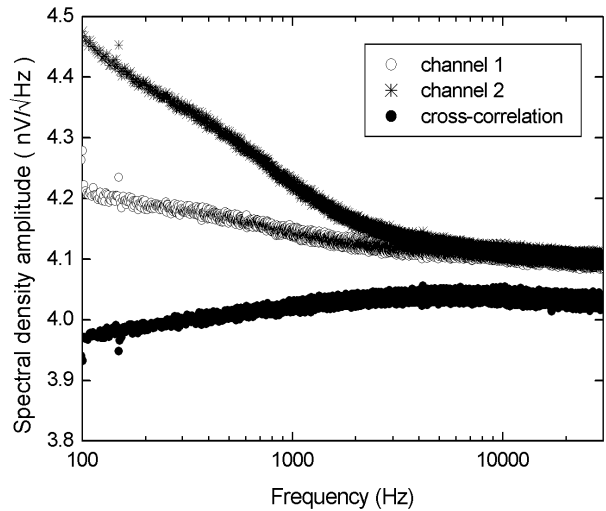}	
	\caption{Amplitude spectra of the cross-correlation signal $N(f_k)$ and of the auto-correlation of the two spectrum analyzer acquisition channels separately. The flicker noise of A is apparent in the autospectra. The shape of $N(f_k)$ is slightly convex because of the bandpass filtering in the second stage of A.}
	\label{fig:spettrorumore}
\end{figure}

A typical spectrum of $N(f_k)$ (averaged over $j$), and the corresponding autospectra of the two channels (rescaled with the same $G_j$), are shown in Fig. \ref{fig:spettrorumore}.

For this experimental run, the relative deviations $\delta^j$ have a mean $\delta =$ \SI{-33}{\micro K  \, K^{-1}}  with a standard deviation $\sigma_\delta =$ \SI{40}{\micro K \, K^{-1}}. Such experimental value for $\sigma_\delta$ is near ($+24\,\%$) the theoretical prediction  for the Type A uncertainty of a measurement performed with an ideal absolute thermometer\footnote{That is, having noiseless amplifiers and an \emph{a priori} known and stable gain.} measuring over the same bandwidth $B$ for same total time $\tau$. More interestingly, $\sigma_\delta$ is somewhat \emph{lower} ($-14\,\%$) than the theoretical prediction for an ideal \emph{relative} noise thermometer, if $\tau$ includes the time required for calibration with a known reference temperature; see Ref. \cite{White2008b} for an extensive explanation. 

\begin{figure}[ht]
	\centering
	\includegraphics[width=3.5in]{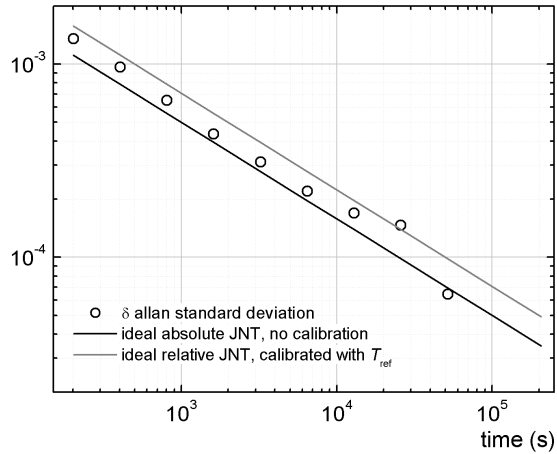}	
	\caption{Allan standard deviation of the JNT error $\delta_j = \Delta T^j/T_{90}$, compared with the corresponding theoretical predictions for an ideal absolute thermometer and an ideal relative thermometer calibrated against a reference temperature $T_\mathrm{ref}$.}
	\label{fig:errore_allan}
\end{figure}

Fig. \ref{fig:errore_allan} shows these same conclusions in graphical form, but when Allan standard deviations are compared; the noise of $\delta$ is still approximately white, therefore the JNT Type A uncertainty is at the moment limited only by the acquisition time. The time for a continuous measurement run is in turn limited by battery charge (the run here presented is close to this limit) but the results of several runs can be averaged together.

An interesting performance test \cite{Zhang2008} is the measurement, instead of the noise $e_\mathrm{N}$ of the resistor $R$, of two independent noises $e_\mathrm{N1}$ and $e_\mathrm{N2}$ generated by two resistors $R_1$ and $R_2$, mounted in a probe having electrical properties similar to that of $R$. An ideal spectrum analyzer outcome would be a null spectrum at any frequency; the spectrum measured gives the magnitude of the systematic errors due to undesired residual correlation effects. In the present setup, the residual correlation has a quadratic frequency dependence, which integrated on $B$ give a systematic deviation lower than \SI{10}{\micro K  \, K^{-1}}. However, the test does not rule out the existence of all residual correlations \cite{Storm86, White84,  White2000} that could occur when a single resistor is measured. 

Consistency tests for different run parameters (bandwidth $B$ extension, sampling rate, $R$ value, $e_\mathrm{C}/e_\mathrm{N}$ ratio, choice of $m_\mathrm{N}$ and $m_\mathrm{C}$, etc) require great experimentation effort, and will be conducted in a systematic way on a future version of the thermometer (see Sec. \ref{sec:perspectives}) for which we expect improved performances. 

\section{Uncertainty}

The JNT uncertainty estimation includes a large number of terms. While some of them are related to well-established instrument properties and calibration techniques (thermometry, resistance and ac voltage measurement), others are matter of careful theoretical evaluations and ad-hoc experiments on acquisition chain of the thermometer \cite{White84,Storm89} and in particular of the properties of the input amplifiers \cite{White2000}. Even the data processing algorithms may cause unexpected systematic errors, as has been very recently pointed out \cite{White2008b}. We didn't yet go through those hard tasks. Therefore, the uncertainty budget given in Tab. \ref{tab:uncertainty} for the relative reading error $\delta$ is intended only as a tool to identify further goals for improvement: some contributions are truly related to our JNT, others are simply taken from literature.

If the estimation of Tab. \ref{tab:uncertainty} is provisionally trusted, the relative difference between the JNT and resistance thermometer readings $\delta$ has an uncertainty $u(\delta) = \SI{58}{\micro K  \, K^{-1}}$, which main contribution come from the JNT reading. Such result can be compared with the recent estimate $u(\delta) =$ \SI{25}{\micro K  \, K^{-1}} of NIST absolute JNT at TPW \cite{Benz2009}, or with those of a number of (relative) acoustic thermometery results \cite{Benedetto2004,Ripple2007}, which consistently estimate $\delta$=\SI{12}{\micro K  \, K^{-1}} with a relative uncertainty as low as \SI{2}{\micro K  \, K^{-1}} near the gallium fixed point ($\approx$\SI{303}{K}).

\begin{table}[ht]
	\centering
	\caption{Uncertainty budget for the thermometer relative error $\delta$.}
	\includegraphics[width=5.5in]{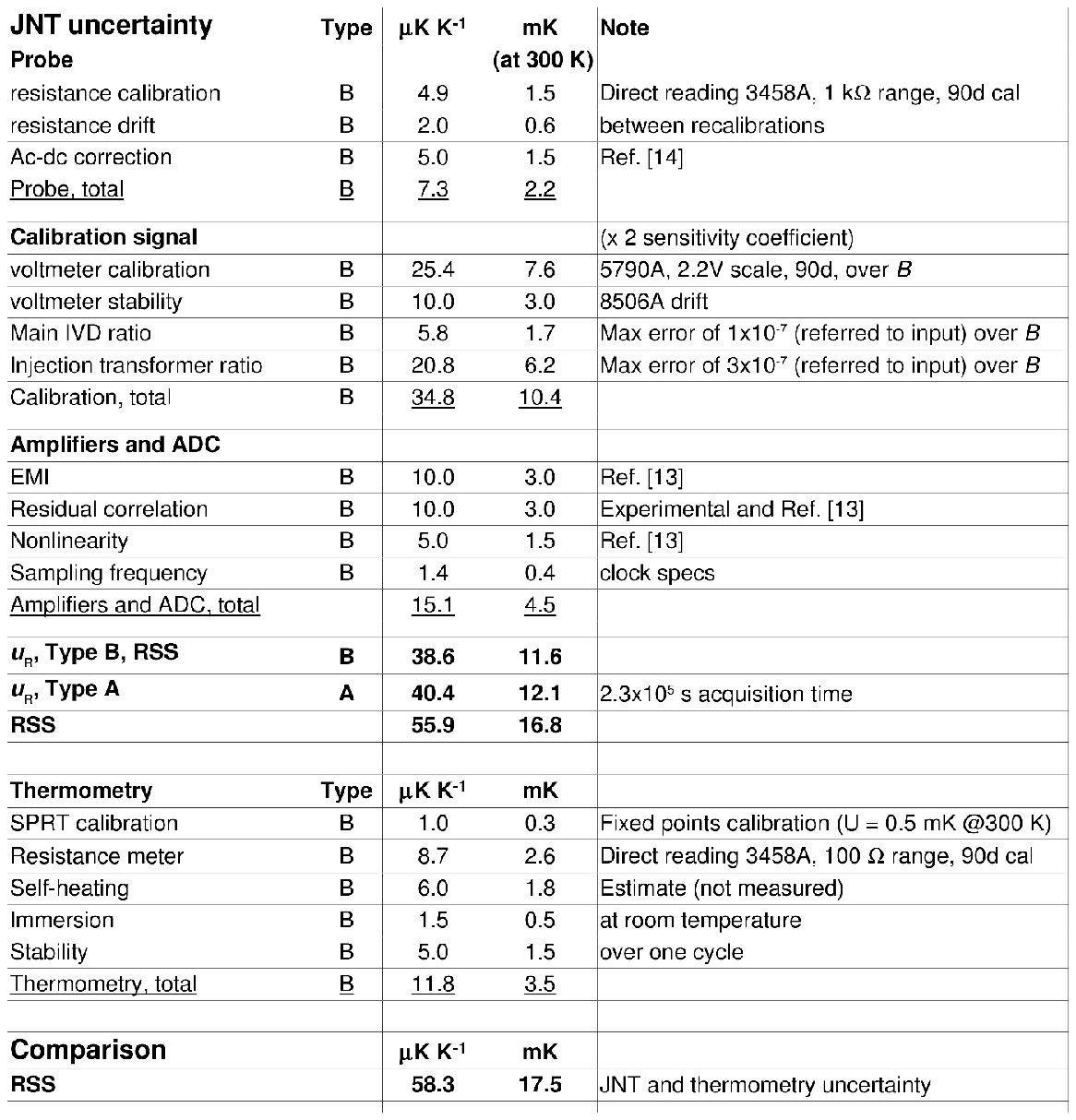}		
	\label{tab:uncertainty}
\end{table}

\section{Conclusion and perspectives}
\label{sec:perspectives}
Looking at Tab. \ref{tab:uncertainty}, we see that a number of uncertainty contributions come from specifications of the commercial instruments employed or, more generally, from measurements for which primary metrology know-how provide better solutions. Therefore, there is room for improvements. We are working on:

\begin{itemize}
\item the development of a thermostat to perform measurements by varying $T$ over a range that includes $T_\mathrm{TPW}$ \cite{Merlone2007};
\item improvements in thermometry measurement setup with the implementation of a dc resistance bridge;
\item the calibration of dividers over $B$ under loading condition;
\item improvements in the measurement of $V^j$ with an automated ac-dc transfer measurement system based on a multijunction thermal converters  \cite{Pogliano2009};
\item the increase in the measurement bandwidth to $B \approx $\SI{20}{kHz};
\item larger battery packs, for an extended measurement time.
\end{itemize}

With these improvements, the measurement uncertainty should drop to the level of \SI{10}{\micro K  \, K^{-1}}; a measurement at $T_\mathrm{TPW}$ will permit the determination of $k_\mathrm{B}$ with the same uncertainty.

\ack
The authors are indebted, for their help in the development of the experiment and fruitful discussions, with their INRIM colleagues R. Gavioso, E. Massa, F. Manta, and A. Merlone; and to M. Ortolano (Politecnico di Torino, Italy). 

\section*{References}

\bibliographystyle{unsrt}
\bibliography{PaperBoltzmann}

\begin{thebibliography}{10}

\bibitem{Johnson28}
J.~B. Johnson.
\newblock Thermal agitation of electricity in conductors.
\newblock {\em Phys. Rev.}, 32:97--109, July 1928.

\bibitem{CCT2005}
2005 {R}ecommendation {T2}: New determinations of thermodynamic temperature and
  the {Boltzmann} constant.
\newblock In {\em Working documents of the 23rd Meeting of the Consultative
  Committee for Thermometry}, 2005.
\newblock {BIPM} document {CCT}/05-31.

\bibitem{Storm86}
L.~Storm.
\newblock Precision measurements of the {Boltzmann} constant.
\newblock {\em Metrologia}, 22:229--234, 1986.

\bibitem{Fellmuth2006}
B.~Fellmuth, Ch. Gaiser, and J.~Fischer.
\newblock Determination of the {B}oltzmann constant -- status and prospects.
\newblock {\em Meas. Sci. Technol.}, 17:R145--R149, 2006.

\bibitem{CODATA2006_RevModPhys}
P.~J. Mohr, B.~N. Taylor, and D.~B. Newell.
\newblock {CODATA} recommended values of the fundamental physical constants:
  2006.
\newblock {\em Rev. Mod. Phys.}, 80:633--730, 2008.

\bibitem{CODATA2006_JPhysChem}
P.~J. Mohr, B.~N. Taylor, and D.~B. Newell.
\newblock {CODATA} recommended values of the fundamental physical constants:
  2006.
\newblock {\em J. Phys. Chem. Ref. Data}, 37:1187--1284, 2008.

\bibitem{White96}
D.~R. White, R.~Galleano, A.~Actis, H.~Brixy, M.~De Groot, J.~Dubbeldam, A.~L.
  Reesink, F.~Edler, H~Sakurai, R.~L. Shepard, and J.~C. Gallop.
\newblock The status of {Johnson} noise thermometry.
\newblock {\em Metrologia}, 33:325--335, 1996.

\bibitem{White2008b}
D.~R. White, S.~P. Benz, J.~R. Labenski, S.~W. Nam, J.~F. Qu, H.~Rogalla, and
  W.~L. Tew.
\newblock Measurement time and statistics for a noise thermometer with a
  synthetic-noise reference.
\newblock {\em Metrologia}, 45:395--405, 2008.

\bibitem{White84}
D.~R. White.
\newblock Systematic errors in a high-accuracy {Johnson} noise thermometer.
\newblock {\em Metrologia}, 20:1--9, 1984.

\bibitem{Storm89}
L.~Storm.
\newblock Errors associated with the measurement of power spectra by the
  correlation technique.
\newblock In {\em Noise in physical systems, including 1/f noise, biological
  systems and membranes: 10th Intl. Conf. proc.}, pages 551--560, Budapest,
  Hungary, 21-25 Aug 1989.

\bibitem{White2000}
D.~R. White and E.~Zimmermann.
\newblock Preamplifier limitations on the accuracy of {Johnson} noise
  thermometers.
\newblock {\em Metrologia}, 37:11--23, 2000.

\bibitem{Nam2003}
Sae~Woo Nam, S.~P. Benz, P.~D. Dresselhaus, W.~L. Tew, D.~R. White, and J.~M.
  Martinis.
\newblock {Johnson} noise thermometry measurements using a quantized voltage
  noise source for calibration.
\newblock {\em {IEEE} Trans. Instr. Meas.}, 52:550--554, April 2003.

\bibitem{White2008}
D.~R. White and S.~P. Benz.
\newblock Constraints on a synthetic noise source for {J}ohnson noise
  thermometry.
\newblock {\em Metrologia}, 45:93--101, 2008.

\bibitem{Benz2009}
S.~P. Benz, J.~Qu, H.~Rogalla, D.~R. White, P.~D. Dresselhaus, W.~L. Tew, and
  S.~W. Nam.
\newblock Improvements in the {NIST} {Johnson} noise thermometry system.
\newblock {\em {IEEE} Trans. Instr. Meas.}
\newblock in press.

\bibitem{Bohacek2006}
J.~Bohacek.
\newblock Reference resistors for calibration of wideband {LCR} meters.
\newblock In {\em Proc. of XVIII IMEKO World Congress: Metrology for a
  sustainable development}, page 00594, Rio de Janeiro, Brazil, 17-22 Sep 2006.

\bibitem{Callegaro08112866}
L.~Callegaro, M.~Pisani, and A.~Pollarolo.
\newblock Very simple {FET} amplifier with a voltage noise floor less than 1
  n{V}/$\sqrt{}${Hz}.
\newblock arXiv:0811.2866 [physics.inst-det].

\bibitem{AB008}
R.~M. Stitt.
\newblock {AC} coupling instrumentation and difference amplifiers, Aug 1991.
\newblock AB-008A Burr-Brown Application Bulletin.

\bibitem{Simonson96}
P.~Simonson and K.-E. Rydler.
\newblock Loading errors in low voltage ac measurements.
\newblock In {\em Precision Electromagnetic Measurements Conf. Digest CPEM'96},
  pages 572--573, Braunschweig, Germany, 17-20 Jun 1996.

\bibitem{Marcarino2003}
P.~Marcarino, P.~P.~M. Steur, and R.~Dematteis.
\newblock Realization at {IMGC} of the {ITS-90} fixed points from the argon
  triple point upwards.
\newblock In {\em Temperature: its measurement and control in science and
  industry, Vol. VII, eighth temperature symposium, AIP. Conf. Proc.}, volume
  684, 29 Sep 2003.

\bibitem{Zhang2008}
J.~T. Zhang and S.~Q. Xue.
\newblock Investigation of the imperfection effect of correlation on {Johnson}
  noise thermometry.
\newblock {\em Metrologia}, 45:436--441, 2008.

\bibitem{Benedetto2004}
G.~Benedetto, R.~M. Gavioso, R.~Spagnolo, P.~Marcarino, and A.~Merlone.
\newblock Re-configurable unit for precise {RMS} measurements.
\newblock {\em {IEEE} Trans. Instr. Meas.}
\newblock in press.

\bibitem{Ripple2007}
D.~C. Ripple, G.~F. Strouse, and M.~R. Moldover.
\newblock Acoustic thermometry results from 271 to 552 {K}.
\newblock {\em Int. J. Thermophys.}, 28:1789--1799, Dec 2007.

\bibitem{Merlone2007}
A.~Merlone, L.~Iacomini, A.~Tiziani, and P.~Marcarino.
\newblock A liquid bath for accurate temperature measurements.
\newblock {\em Measurement}, 40:422--427, 2007.

\bibitem{Pogliano2009}
U.~Pogliano, B.~Trinchera, and F.~Francone.
\newblock Re-configurable unit for precise {RMS} measurements.
\newblock {\em {IEEE} Trans. Instr. Meas.}
\newblock in press.

\end{thebibliography}

\end{document}